\documentclass[12pt]{article}
\usepackage{amsfonts}
\begin{document}
\begin{center}
{\Large\bf Gauge symmetries of strings in supertwistor space}\\[0.5cm]
{\large D.V.~Uvarov\footnote{E-mail: d\_uvarov@hotmail.com, uvarov@kipt.kharkov.ua}}\\[0.4cm]
{\it NSC Kharkov Institute of Physics and Technology,}\\ {\it 61108 Kharkov, Ukraine}\\[0.5cm]
\end{center}
\begin{abstract}
Recently we have considered supertwistor reformulation of the $D=4$ $N=1,2$ superstring action that comprises
Newman-Penrose dyad components and is classically equivalent to
the Green-Schwarz one. It was shown that in the covariant
$\kappa-$symmetry gauge the supertwistor representation of the
string action simplifies. Here we analyze its Hamiltonian
formulation, classify the constraints on the phase-space
variables, and find the covariant set of generators of the
gauge symmetries. Quantum symmetries of the supertwistor
representation of the string action are examined by applying the
world-sheet CFT technique. Considered are various generalizations
of the model from the perspective of their possible relation to
known twistor superstring models.
\end{abstract}

\section{Introduction}
Application of the (super)twistor methods \cite{twistor},
\cite{Ferber} sheds new light not only on the Yang-Mills theory
and gravity \cite{WW} but also on the features of point-like
\cite{Ferber}, \cite{Shirafuji}-\cite{Bette} and extended
(super)symmetric objects \cite{Ilyenko}-\cite{FedLuk}. Not long ago
this statement have got further evidence by constructing 
the topological string model \cite{Witten} in
$\mathbb{CP}^{(3|4)}$ projective supertwistor space that it is
related to the perturbative sector of conformal $\mathcal N=4$ SYM
theory. In a sense this correspondence can be viewed as the weak
coupling regime counterpart of the AdS/CFT-correspondence
\cite{AdS/CFT} that is presently best understood in the
supergravity limit of $IIB$ string theory dual to the strong
coupling regime of the $\mathcal N=4$ SYM. Witten's insight
fostered progress in investigation of the perturbative YM theory, geometry of
superCalabi-Yau manifolds and string theory (for review see
\cite{Trieste}). In particular there were proposed nontopological
twistor superstring models \cite{Berkovits}-\cite{Bars}.
Nontopological twistor superstring model \cite{Berkovits},
\cite{BerkovitsMotl} has been shown to provide another way of
computation of the tree level $\mathcal N=4$ SYM scattering amplitudes that
initially have got stringy interpretation in \cite{Witten}.

Motivated by the above results stemming from the synthesis of twistor theory and strings we have considered \cite{CQG06} the reformulation of the superstring in $D=4$ $N=1,2$ superspace in terms of supertwistors. We have started with the Lorentz-harmonic \cite{GIKOS}-\cite{harmonics} formulation of the superstring action proposed and studied in \cite{BZstring}
\begin{equation}\label{1}
S=S_{kin}+S_{WZ},
\end{equation}
where
\begin{equation}\label{1k}
S_{kin}=\int\limits_{{\rm M}^2}d^2\xi e(\xi)\left(-{\textstyle\frac{1}{2(\alpha^\prime)^{1/2}}}[e^{\mu+2}(\xi)n^{-2}_m(\xi)+e^{\mu-2}(\xi)n_m^{+2}(\xi)]\omega^m_\mu(\xi)+c\right)
\end{equation}
is the Lorentz-harmonic representation of the superstring kinetic term and the Wess-Zumino term has the standard form
\begin{equation}\label{1w}
\begin{array}{rl}
S_{WZ}=&\frac{is}{c\alpha^\prime}{\displaystyle\int\limits_{{\rm M}^2}}d^2\xi\varepsilon^{\mu\nu}\omega^{\dot\alpha\alpha}_\mu(\partial_\nu\theta^1_\alpha\bar\theta^1_{\dot\alpha}-\theta^1_\alpha \partial_\nu\bar\theta^1_{\dot\alpha}-\partial_\nu\theta^2_\alpha\bar\theta^2_{\dot\alpha}+\theta^2_\alpha \partial_\nu\bar\theta^2_{\dot\alpha})\\[0.2cm]
+&\frac{2s}{c\alpha^\prime}{\displaystyle\int\limits_{{\rm M}^2}}d^2\xi\varepsilon^{\mu\nu}(\partial_\mu\theta^{1\alpha}\bar\theta^{1\dot\alpha}-\theta^{1\alpha}\partial_\mu\bar\theta^{1\dot\alpha})(\partial_\nu\theta^2_\alpha\bar\theta^2_{\dot\alpha}-\theta^2_\alpha \partial_\nu\bar\theta^2_{\dot\alpha}).
\end{array}
\end{equation}
In the above formulas $\xi^\mu=(\tau,\sigma)$ are world-sheet local coordinates; $\omega^m_\mu=\partial_\mu x^m+i\partial_\mu\theta^{I\alpha}\sigma^m_{\alpha\dot\alpha}\bar\theta^{I\dot\alpha}-i\theta^{I\alpha}\sigma^m_{\alpha\dot\alpha}\partial_\mu\bar\theta^{I\dot\alpha}$ and $\partial_\mu\theta^{I\alpha}$, $\partial_\mu\bar\theta^{I\dot\alpha}$ ($I=1,2$) are the pullbacks onto the world sheet of target-space supersymmetric 1-forms; $e^{\pm2}_\mu=e^0_\mu\pm e^1_\mu$ and $e^{\mu\pm2}=e^{\mu0}\pm e^{\mu1}$ are auxiliary zweibein $e^f_\mu$ and its inverse $e^\mu_f$ components written in the $2d$ light-cone basis characterized by $\pm2$ weights w.r.t. world-sheet Lorentz group $SO(1,1)$, $e=det(e^{f}_\mu)$; $n^{\pm2}_m(\xi)$ are auxiliary light-like vectors from the Cartan-Penrose space-time repere attached to the world sheet, that can be normalized as $n^{+2}_m\eta^{mn}n^{-2}_n=2$; $s=\pm1$ is the numerical factor; $\alpha'$ is the Regge slope parameter so that the string tension is $T=\frac{1}{2c\alpha'}$. Variation of the action (\ref{1}) w.r.t. auxiliary variables allows to express the pullback of supersymmetric  1-form $\omega^m_\mu$ in terms of them as
\begin{equation}\label{2}
\omega^m_\mu={\textstyle\frac{c\alpha'}{2}}(e_\mu^{+2}n^{m-2}+e_\mu^{-2}n^{m+2})
\end{equation}
providing the Lorentz-covariant resolution of the Virasoro constraints. Expression (\ref{2}) can be used to exclude auxiliary vectors $n^{\pm2}_m$ from the action (\ref{1}) to establish its classical equivalence to the GS action. Note that the above light-like vectors from the Cartan-Penrose moving repere allow the following realization 
\begin{equation}
n^{+2}_m=v^{\alpha+}\sigma_{m\alpha\dot\alpha}\bar v^{\dot\alpha+},\quad n^{-2}_m=u^{\alpha-}\sigma_{m\alpha\dot\alpha}\bar u^{\dot\alpha-}
\end{equation}
in terms of two-component $SL(2,\mathbb C)$ spinors that admit interpretation as Newman-Penrose dyad components subject to normalization $u^\alpha v_\alpha=\bar u^{\dot\alpha}\bar v_{\dot\alpha}=1$\footnote{Below to avoid complication of notation we skip $SO(1,1)$ weights of auxiliary spinors.}. It is used to ensure invariance of the action (\ref{1}) under irreducible $\kappa$-symmetry transformations. 

In \cite{CQG06} it was shown that (\ref{1}) can be converted into the supertwistor form by introducing two supertwistors subject to four constraints to maintain the correspondence with the original superspace-time description. Twistorization of the kinetic term is analogous to the superparticle \cite{Shirafuji}-\cite{Bette} and tensionless super $p$-brane \cite{Ilyenko}-\cite{BdAM}, while auxiliary spinors are introduced into the WZ term via the completeness relations $u^\alpha v_\beta-v^\alpha u_\beta=\delta^\alpha_\beta$ and c.c. The resulting supertwistor action was shown to be invariant under $\kappa$-symmetry. It was also established \cite{CQG06} that upon covariant $\kappa$-symmetry gauge fixation the supertwistor action reduces to the quadratic one
\begin{equation}\label{3}
S_{tw}=-\!\int\limits_{\rm M^2}\!\! d^2\xi{\textstyle\frac{ie}{4(\alpha')^{1/2}}}\left(e^{\mu+2}(\partial_\mu\bar{\mathbb Z}_{\mathbb A}\mathbb Z^{\mathbb A}-\bar{\mathbb Z}_{\mathbb A}\partial_\mu\mathbb Z^{\mathbb A})\!+\! e^{\mu-2}(\partial_\mu\bar{\mathbb W}_{\mathbb A}\mathbb W^{\mathbb A}-\bar{\mathbb W}_{\mathbb A}\partial_\mu\mathbb W^{\mathbb A})\right)+c\!\int\limits_{\rm M^2}\!\! d^2\xi e,
\end{equation}
where $\mathbb Z^{\mathbb A}$ can either be the twistor or supertwistor and similarly $\mathbb W^{\mathbb A}$. So that the action (\ref{3}) uniformly describes bosonic string (both $\mathbb Z^{\mathbb A}$ and $\mathbb W^{\mathbb A}$ are twistors), gauge fixed $N=1$ heterotic superstring ($\mathbb Z^{\mathbb A}$ is the supertwistor and $\mathbb W^{\mathbb A}$ is the twistor or vice versa depending on the sign $s$ of the WZ term) and $N=2$ superstring (both $\mathbb Z^{\mathbb A}$ and $\mathbb W^{\mathbb A}$ are supertwistors).

Here we examine in detail the Hamiltonian mechanics and symmetries of the $N=2$ superstring. Particular results for other models can be derived from those to be presented. Let us note that the presence of the cosmological term $\int d^2\xi e(\xi)=\int d^2\xi\sqrt{-g(\xi)}$ spoils the $2d$ Weyl invariance of the action (\ref{1}), (\ref{3}) and prevents all of the auxiliary zweibein components to be gauged away so we would like to consider its Weyl-invariant generalization
\begin{equation}
\int\limits_{\rm M^2}d^2\xi e(\xi)\rightarrow\int\limits_{\rm M^2}d^2\xi e(\xi)n(\xi)\bar n(\xi),
\end{equation}
where $n=u^\alpha v_\alpha$, $\bar n=\bar u^{\dot\alpha}\bar v_{\dot\alpha}$.
Such modification amounts to introducing unnormalized Newman-Penrose dyad instead of the normalized one obeying $n=\bar n=1$. It does not increase the number of degrees of freedom since one trades two second-class constraints $\Xi=n-1\approx0$, $\bar\Xi=\bar n-1\approx0$ for a single first-class constraint - the Weyl symmetry generator.
The action (\ref{3}) has to be supplemented by terms that enforce zero norm constraints on (super)twistors
\begin{equation}
\chi_{\mathbb Z}=\bar{\mathbb Z}_{\mathbb A}\mathbb Z^{\mathbb A}=0,\quad\chi_{\mathbb W}=\bar{\mathbb W}_{\mathbb A}\mathbb W^{\mathbb A}=0,
\end{equation}
and also 2 orthogonality constraints
\begin{equation}
\chi_{\bar{\mathbb W}\mathbb Z}=\bar{\mathbb W}_{\mathbb A}\mathbb Z^{\mathbb A}=0,\quad\chi_{\bar{\mathbb Z}\mathbb W}= \bar{\mathbb Z}_{\mathbb A}\mathbb W^{\mathbb A}=0
\end{equation}
via the Lagrange multipliers
\begin{equation}
S_{aux}=\int\limits_{\rm M^2}d^2\xi(s_{\mathbb Z}\chi_{\mathbb Z}+s_{\mathbb W}\chi_{\mathbb W}+s_{\bar{\mathbb W}\mathbb Z}\chi_{\bar{\mathbb W}\mathbb Z}+s_{\bar{\mathbb Z}\mathbb W}\chi_{\bar{\mathbb Z}\mathbb W}).
\end{equation}
These constraints ensure that the matrix $x^{\dot\alpha\alpha}=\tilde\sigma^{\dot\alpha\alpha}_mx^m$ is Hermitean, i.e. superspace bosonic body is real.
Lagrange multipliers $s_{\mathbb Z}$ and $s_{\mathbb W}$ that are
real and independent in our case admit interpretation as world-sheet $U(1)$ gauge fields.

\section{Hamiltonian mechanics of $D=4$ superstrings in supertwistor formulation}

Thus we consider the $\kappa-$symmetry gauge-fixed $N=2$ superstring model characterized by the following Lagrangian density\footnote{This Lagrangian density can be derived upon $\kappa$-symmetry gauge fixing the $N=2$ superstring by the conditions $\theta^{1\alpha}=\theta^{2\alpha}$, $\bar\theta^{1\dot\alpha}=\bar\theta^{2\dot\alpha}$ as was found in \cite{CQG06}.}
\begin{equation}\label{4}
\begin{array}{rl}
L(\xi)=&-\frac{i}{4c\alpha'}\left(\rho^{\mu}(\partial_\mu\bar{\mathcal Z}_{A}\mathcal Z^{A}-\bar{\mathcal Z}_{A}\partial_\mu\mathcal Z^{A})\!+\! \varrho^{\mu}(\partial_\mu\bar{\mathcal W}_{A}\mathcal W^{A}-\bar{\mathcal W}_{A}\partial_\mu\mathcal W^{A})\right)
\\[0.2cm]
&+\frac{1}{2c\alpha'}\varepsilon_{\mu\nu}\varrho^{\mu}\rho^{\nu}n\bar n,
\end{array}
\end{equation}
where $\rho^{\mu}=(\alpha')^{1/2}ee^{\mu+2}$, $\varrho^{\mu}=(\alpha')^{1/2}ee^{\mu-2}$ are dimensionless zweibein density components that are more convenient to deal with in the Hamiltonian formulation \cite{BZstring}. The supertwistor world-sheet variables $\mathcal Z^{\mathcal A}(\tau,\sigma)$, $\bar{\mathcal Z}_{\mathcal A}(\tau,\sigma)$ and $\mathcal W^A(\tau,\sigma)$, $\bar{\mathcal W}_A(\tau,\sigma)$ have the following dependence on $D=4$ $N=1$ superspace coordinates $(x^{\dot\alpha\alpha}, \theta^\alpha, \bar\theta^{\dot\alpha})$
\begin{equation}
\begin{array}{c}
\mathcal Z^{A}=(\mu^\alpha, \bar u_{\dot\alpha}, \bar\eta):\quad\mu^\alpha=i(x^{\dot\alpha\alpha}+2i\theta^\alpha\bar\theta^{\dot\alpha})\bar u_{\dot\alpha},\quad\bar\eta=2\bar u^{\dot\alpha}\bar\theta_{\dot\alpha};\\[0.2cm]
\bar{\mathcal Z}_{A}=(u_\alpha, \bar\mu^{\dot\alpha}, \eta):\quad\bar\mu^{\dot\alpha}=-i(x^{\dot\alpha\alpha}-2i\theta^\alpha\bar\theta^{\dot\alpha})u_\alpha,\quad\eta=2u^\alpha\theta_\alpha,
\end{array}
\end{equation}
and
\begin{equation}
\begin{array}{c}
\mathcal W^A=(\nu^\alpha, \bar v_{\dot\alpha}, \bar\zeta):\quad\nu^\alpha=i(x^{\dot\alpha\alpha}+2i\theta^\alpha\bar\theta^{\dot\alpha})\bar v_{\dot\alpha},\quad\bar\zeta=2\bar v^{\dot\alpha}\bar\theta_{\dot\alpha};\\[0.2cm]
\quad\bar{\mathcal W}_A=(v_\alpha, \bar\nu^{\dot\alpha}, \zeta):\quad\bar\nu^{\dot\alpha}=-i(x^{\dot\alpha\alpha}-2i\theta^\alpha\bar\theta^{\dot\alpha})v_\alpha,\quad\zeta=2v^\alpha\theta_\alpha.
\end{array}
\end{equation}
 Lagrangian density (\ref{4}) can be split into two parts, one of which contains solely $\tau$-derivatives
\begin{equation}\label{5}
L(\tau,\sigma)=L_\tau+L_\sigma,
\end{equation}
where
\begin{equation}
\begin{array}{rl}
L_\tau=&-\frac{i}{4c\alpha'}\rho^{\tau}(\partial_\tau u_\alpha\mu^\alpha+\partial_\tau{\bar\mu}{}^{\dot\alpha}\bar u_{\dot\alpha}+\partial_\tau\eta\bar\eta-u_\alpha\partial_\tau\mu^\alpha-\bar\mu^{\dot\alpha}\partial_\tau{\bar u}_{\dot\alpha}-\eta\partial_\tau{\bar\eta})\\[0.2cm]
-&\frac{i}{4c\alpha'}\varrho^{\tau}(\partial_\tau v_\alpha\nu^\alpha+\partial_\tau{\bar\nu}{}^{\dot\alpha}\bar v_{\dot\alpha}+\partial_\tau\zeta\bar\zeta-v_\alpha\partial_\tau\nu^\alpha-\bar\nu^{\dot\alpha}\partial_\tau{\bar v}_{\dot\alpha}-\zeta\partial_\tau\bar\zeta),
\end{array}
\end{equation}
\begin{equation}
\begin{array}{rl}
L_\sigma=&
-\frac{i}{4c\alpha'}\rho^{\sigma}(\partial_\sigma u_\alpha\mu^\alpha+\partial_\sigma{\bar\mu}{}^{\dot\alpha}\bar u_{\dot\alpha}+\partial_\sigma\eta\bar\eta-u_\alpha\partial_\sigma\mu^\alpha-\bar\mu^{\dot\alpha}\partial_\sigma{\bar u}_{\dot\alpha}-\eta\partial_\sigma{\bar\eta})\\[0.2cm]
-&\frac{i}{4c\alpha'}\varrho^{\sigma}(\partial_\sigma v_\alpha\nu^\alpha+\partial_\sigma{\bar\nu}{}^{\dot\alpha}\bar v_{\dot\alpha}+\partial_\sigma\zeta\bar\zeta-v_\alpha\partial_\sigma\nu^\alpha-\bar\nu^{\dot\alpha}\partial_\sigma{\bar v}_{\dot\alpha}-\zeta\partial_\sigma\bar\zeta)\\[0.2cm]
-&\frac{1}{2c\alpha'}n\bar n(\varrho^{\tau}\rho^{\sigma}-\rho^{\tau}\varrho^{\sigma}).
\end{array}
\end{equation}

Definition of the canonical momenta densities
\begin{equation}
\mathfrak P^{\mathfrak M}(\tau,\sigma)=(p_{(\mu)\alpha}, \bar p_{(\mu)\dot\alpha}, p^\alpha_{(u)}, \bar p^{\dot\alpha}_{(u)}, p_{(\nu)\alpha}, \bar p_{(\nu)\dot\alpha}, p^\alpha_{(v)}, \bar p^{\dot\alpha}_{(v)}, p_{(\eta)}, \bar p_{(\eta)}, p_{(\zeta)}, \bar p_{(\zeta)}, P_\mu^{(\rho)}, P_\mu^{(\varrho)})=\frac{\delta L_\tau}{\delta\partial_\tau{\mathfrak Q}}
\end{equation}
conjugate to the supertwistor and zweibein density components
\begin{equation}
\mathfrak Q_{\mathfrak N}(\tau,\sigma)=(\mu^\beta, \bar\mu^{\dot\beta}, u_\beta, \bar u_{\dot\beta}, \nu^\beta, \bar\nu^{\dot\beta}, v_\beta, \bar v_{\dot\beta}, \eta, \bar\eta, \zeta, \bar\zeta, \rho^{\nu}, \varrho^{\nu})
\end{equation}
yields the primary constraints
\begin{equation}\label{10}
\begin{array}{c}
T_{(\mu)\alpha}=p_{(\mu)\alpha}-\frac{i}{4c\alpha'}\rho^{\tau}u_\alpha\approx0,\quad\bar T_{(\mu)\dot\alpha}=\bar p_{(\mu)\dot\alpha}+\frac{i}{4c\alpha'}\rho^{\tau}\bar u_{\dot\alpha}\approx0;\\[0.2cm]
T^\alpha_{(u)}=p^\alpha_{(u)}+\frac{i}{4c\alpha'}\rho^{\tau}\mu^\alpha\approx0,\quad\bar T^{\dot\alpha}_{(u)}=\bar p^{\dot\alpha}_{(u)}-\frac{i}{4c\alpha'}\rho^{\tau}\bar\mu^{\dot\alpha}\approx0;\\[0.2cm]
T_{(\eta)}=p_{(\eta)}+\frac{i}{4c\alpha'}\rho^{\tau}\bar\eta\approx0,\quad\bar T_{(\eta)}=\bar p_{(\eta)}+\frac{i}{4c\alpha'}\rho^{\tau}\eta\approx0;
\end{array}
\end{equation}
\begin{equation}\label{10'}
P_\mu^{(\rho)}\approx0,
\end{equation}
where $\eta^\ast=\bar\eta$ and $(p_{(\eta)})^\ast=-\bar p_{(\eta)}$, and similarly for the $\mathcal W$-supertwistor sector variables
\begin{equation}\label{11}
\begin{array}{c}
T_{(\nu)\alpha}=p_{(\nu)\alpha}-\frac{i}{4c\alpha'}\varrho^{\tau}v_\alpha\approx0,\quad\bar T_{(\nu)\dot\alpha}=\bar p_{(\nu)\dot\alpha}+\frac{i}{4c\alpha'}\varrho^{\tau}\bar v_{\dot\alpha}\approx0;\\[0.2cm]
T^\alpha_{(v)}=p^\alpha_{(v)}+\frac{i}{4c\alpha'}\varrho^{\tau}\nu^\alpha\approx0,\quad\bar T^{\dot\alpha}_{(v)}=\bar p^{\dot\alpha}_{(v)}-\frac{i}{4c\alpha'}\varrho^{\tau}\bar\nu^{\dot\alpha}\approx0;\\[0.2cm]
T_{(\zeta)}=p_{(\zeta)}+\frac{i}{4c\alpha'}\varrho^{\tau}\bar\zeta\approx0,\quad\bar T_{(\zeta)}=\bar p_{(\zeta)}+\frac{i}{4c\alpha'}\varrho^{\tau}\zeta\approx0;
\end{array}
\end{equation}
\begin{equation}\label{11'}
P_\mu^{(\varrho)}\approx0.
\end{equation}
Canonical momenta are defined to have the following Poisson brackets (P.B.) with the conjugate coordinates
\begin{equation}
\{\mathfrak P^{\mathfrak M}(\sigma),\mathfrak Q_{\mathfrak N}(\sigma')\}_{P.B.}=\delta_{\mathfrak N}^{\mathfrak M}\delta(\sigma-\sigma').
\end{equation}
Introduction of the momenta densities allows to present $L_\tau$ part of the Lagrangian (\ref{5}) as
\begin{equation}\begin{array}{rl}
L_\tau\approx& \partial_\tau u_\alpha p^\alpha_{(u)}+\partial_\tau\bar u_{\dot\alpha}\bar p^{\dot\alpha}_{(u)}+\partial_\tau\mu^\alpha p_{(\mu)\alpha}+\partial_\tau\bar\mu^{\dot\alpha}\bar p_{(\mu)\dot\alpha}+\partial_\tau\eta p_{(\eta)}+\partial_\tau\bar\eta\bar p_{(\eta)}\\[0.2cm]
+&\partial_\tau v_\alpha p^\alpha_{(v)}+\partial_\tau\bar v_{\dot\alpha}\bar p^{\dot\alpha}_{(v)}+\partial_\tau\nu^\alpha p_{(\nu)\alpha}+\partial_\tau\bar\nu^{\dot\alpha}\bar p_{(\nu)\dot\alpha}+\partial_\tau\zeta p_{(\zeta)}+\partial_\tau\bar\zeta\bar p_{(\zeta)}.
\end{array}
\end{equation}
Therefore canonical Hamiltonian density equals
\begin{equation}
\begin{array}{c}
H_0(\tau,\sigma)=-L_\sigma\\[0.2cm]
=\frac{i}{4c\alpha'}\rho^{\sigma}(\partial_\sigma u_\alpha\mu^\alpha+\partial_\sigma{\bar\mu}{}^{\dot\alpha}\bar u_{\dot\alpha}+\partial_\sigma\eta\bar\eta-u_\alpha\partial_\sigma\mu^\alpha-\bar\mu^{\dot\alpha}\partial_\sigma{\bar u}_{\dot\alpha}-\eta\partial_\sigma{\bar\eta}-2in\bar n\varrho^{\tau})\\[0.2cm]
+\frac{i}{4c\alpha'}\varrho^{\sigma}(\partial_\sigma v_\alpha\nu^\alpha+\partial_\sigma{\bar\nu}{}^{\dot\alpha}\bar v_{\dot\alpha}+\partial_\sigma\zeta\bar\zeta-v_\alpha\partial_\sigma\nu^\alpha-\bar\nu^{\dot\alpha}\partial_\sigma{\bar v}_{\dot\alpha}-\zeta\partial_\sigma\bar\zeta+2in\bar n\rho^{\tau})
\end{array}
\end{equation}
and the total Hamiltonian density is presented as the sum of $H_0$ and the primary constraints with arbitrary Lagrange multipliers
\begin{equation}
\begin{array}{rl}
H_T(\tau,\sigma)=& H_0+a^\alpha T_{(\mu)\alpha}+\bar a^{\dot\alpha}\bar T_{(\mu)\dot\alpha}+b_\alpha T^\alpha_{(u)}+\bar b_{\dot\alpha}\bar T^{\dot\alpha}_{(u)}+\lambda T_{(\eta)}+\bar\lambda\bar T_{(\eta)}\\[0.2cm]
+& c^\alpha T_{(\nu)\alpha}+\bar c^{\dot\alpha}\bar T_{(\nu)\dot\alpha}+d_\alpha T^\alpha_{(v)}+\bar d_{\dot\alpha}\bar T^{\dot\alpha}_{(v)}+\pi T_{(\zeta)}+\bar\pi\bar T_{(\zeta)}\\[0.2cm]
+& s_{\mathcal Z}\chi_{\mathcal Z}+s_{\mathcal W}\chi_{\mathcal W}+s_{\bar{\mathcal W}\mathcal Z}\chi_{\bar{\mathcal W}\mathcal Z}+s_{\bar{\mathcal Z}\mathcal W}\chi_{\bar{\mathcal Z}\mathcal W}+\beta^{\mu(\rho)}P^{(\rho)}_{\mu}+\beta^{\mu(\varrho)}P^{(\varrho)}_{\mu},
\end{array}
\end{equation}
where the zero norm and orthogonality constraints in components read
\begin{equation}\label{6}
\chi_{\mathcal Z}=u_\alpha\mu^\alpha+\bar\mu^{\dot\alpha}\bar u_{\dot\alpha}+\eta\bar\eta\approx0,\quad\chi_{\mathcal W}=v_\alpha\nu^\alpha+\bar\nu^{\dot\alpha}\bar v_{\dot\alpha}+\zeta\bar\zeta\approx0
\end{equation}
and
\begin{equation}\label{7}
\chi_{\bar{\mathcal W}\mathcal Z}=v_\alpha\mu^\alpha+\bar\nu^{\dot\alpha}\bar u_{\dot\alpha}+\zeta\bar\eta\approx0,\quad
\chi_{\bar{\mathcal Z}\mathcal W}=u_\alpha\nu^\alpha+\bar\mu^{\dot\alpha}\bar v_{\dot\alpha}+\eta\bar\zeta\approx0.
\end{equation}
Lagrange multipliers $a$, $b$, $c$, $d$, $s$ and $\beta$ are assumed to be commuting, whereas $\lambda$ and $\pi$ are anticommuting.

Following the Dirac method \cite{Dirac} we have to study the temporal conservation in the weak sense of the constraints (\ref{10})-(\ref{11'}), (\ref{6}), (\ref{7}) w.r.t. evolution generated by the total Hamiltonian $H_T$.
The conservation of the constraints $P^{(\rho,\varrho)}_\sigma\approx0$ yields pair of the secondary constraints
\begin{equation}\label{8}
T_{\mathcal Z\sigma}=\frac{i}{4}(\partial_\sigma u_\alpha\mu^\alpha+\partial_\sigma{\bar\mu}{}^{\dot\alpha}\bar u_{\dot\alpha}+\partial_\sigma\eta\bar\eta-u_\alpha\partial_\sigma\mu^\alpha-\bar\mu^{\dot\alpha}\partial_\sigma{\bar u}_{\dot\alpha}-\eta\partial_\sigma{\bar\eta}-2in\bar n\varrho^{\tau})\approx0,
\end{equation}
and
\begin{equation}\label{9}
T_{\mathcal W\sigma}=\frac{i}{4}(\partial_\sigma v_\alpha\nu^\alpha+\partial_\sigma{\bar\nu}{}^{\dot\alpha}\bar v_{\dot\alpha}+\partial_\sigma\zeta\bar\zeta-v_\alpha\partial_\sigma\nu^\alpha-\bar\nu^{\dot\alpha}\partial_\sigma{\bar v}_{\dot\alpha}-\zeta\partial_\sigma\bar\zeta+2in\bar n\rho^{\tau})\approx0.
\end{equation}
The conservation of the $T_{(\mu)}\approx0$ constraints determines Lagrange multipliers $b$
\begin{equation}
b_\alpha=-{\textstyle\frac{\rho^{\sigma}}{\rho^{\tau}}}\partial_\sigma u_\alpha-\bar k_{\mathcal Z}u_\alpha+{\textstyle\frac{2ic\alpha's_{\bar{\mathcal W}\mathcal Z}}{\rho^{\tau}}}v_\alpha,\quad
\bar b_{\dot\alpha}=-{\textstyle\frac{\rho^{\sigma}}{\rho^{\tau}}}\partial_\sigma\bar u_{\dot\alpha}-k_{\mathcal Z}\bar u_{\dot\alpha}-{\textstyle\frac{2ic\alpha's_{\bar{\mathcal Z}\mathcal W}}{\rho^{\tau}}}\bar v_{\dot\alpha},
\end{equation}
where $k_{\mathcal Z}=(\partial_\sigma\rho^{\sigma}+\beta^{\tau(\rho)}+4ic\alpha's_{\mathcal Z})/(2\rho^{\tau})$,
utilizing that $\rho^{\tau}\not=0$.
The conservation of the $T_{(u)}\approx0$ constraints determines Lagrange multipliers $a$
\begin{equation}
a^\alpha=-{\textstyle\frac{\rho^{\sigma}}{\rho^{\tau}}}\partial_\sigma\mu^\alpha-k_{\mathcal Z}\mu^\alpha-{\textstyle\frac{2ic\alpha' s_{\bar{\mathcal Z}\mathcal W}}{\rho^{\tau}}}\nu^\alpha+{\textstyle\frac{i\bar n R}{\rho^{\tau}}}v^\alpha,\quad
\bar a^{\dot\alpha}=-{\textstyle\frac{\rho^{\sigma}}{\rho^{\tau}}}\partial_\sigma\bar\mu^{\dot\alpha}-\bar k_{\mathcal Z}\bar\mu^{\dot\alpha}+{\textstyle\frac{2ic\alpha' s_{\bar{\mathcal W}\mathcal Z}}{\rho^{\tau}}}\bar\nu^{\dot\alpha}-{\textstyle\frac{inR}{\rho^{\tau}}}\bar v^{\dot\alpha},
\end{equation}
where $R=\rho^{\sigma}\varrho^{\tau}-\varrho^{\sigma}\rho^{\tau}$.
The conservation of the $T_{(\nu)}\approx0$ constraints determines Lagrange multipliers $d$
\begin{equation}
d_\alpha=-{\textstyle\frac{\varrho^{\sigma}}{\varrho^{\tau}}}\partial_\sigma v_\alpha-\bar k_{\mathcal W}v_\alpha+{\textstyle\frac{2ic\alpha' s_{\bar{\mathcal Z}\mathcal W}}{\varrho^{\tau}}}u_\alpha,\quad
\bar d_{\dot\alpha}=-{\textstyle\frac{\varrho^{\sigma}}{\varrho^{\tau}}}\partial_\sigma\bar v_{\dot\alpha}-k_{\mathcal W}\bar v_{\dot\alpha}-{\textstyle\frac{2ic\alpha' s_{\bar{\mathcal W}\mathcal Z}}{\varrho^{\tau}}}\bar u_{\dot\alpha},
\end{equation}
where $k_{\mathcal W}=(\partial_\sigma\varrho^{\sigma}+\beta^{\tau(\varrho)}+4ic\alpha's_{\mathcal W})/(2\varrho^{\tau})$,
under the assumption that $\varrho^{\tau}\not=0$.
The conservation of the $T_{(v)}\approx0$ constraints determines Lagrange multipliers $c$
\begin{equation}
c^\alpha=-{\textstyle\frac{\varrho^{\sigma}}{\varrho^{\tau}}}\partial_\sigma\nu^\alpha-k_{\mathcal W}\nu^\alpha-{\textstyle\frac{2ic\alpha' s_{\bar{\mathcal W}\mathcal Z}}{\varrho^{\tau}}}\mu^\alpha-{\textstyle\frac{i\bar n R}{\varrho^{\tau}}}u^\alpha,\quad
\bar c^{\dot\alpha}=-{\textstyle\frac{\varrho^{\sigma}}{\varrho^{\tau}}}\partial_\sigma\bar\nu^{\dot\alpha}-\bar k_{\mathcal W}\bar\nu^{\dot\alpha}+{\textstyle\frac{2ic\alpha' s_{\bar{\mathcal Z}\mathcal W}}{\varrho^{\tau}}}\bar\mu^{\dot\alpha}+{\textstyle\frac{inR}{\varrho^{\tau}}}\bar u^{\dot\alpha}.
\end{equation}
The conservation of fermionic constraints $T_{(\eta)}\approx0$ determines fermionic Lagrange multipliers $\lambda$
\begin{equation}
\lambda=-{\textstyle\frac{\rho^{\sigma}}{\rho^{\tau}}}\partial_\sigma\eta-\bar k_{\mathcal Z}\eta+{\textstyle\frac{2ic\alpha' s_{\bar{\mathcal W}\mathcal Z}}{\rho^{\tau}}}\zeta,\quad
\bar\lambda=-{\textstyle\frac{\rho^{\sigma}}{\rho^{\tau}}}\partial_\sigma\bar\eta-k_{\mathcal Z}\bar\eta-{\textstyle\frac{2ic\alpha' s_{\bar{\mathcal Z}\mathcal W}}{\rho^{\tau}}}\bar\zeta.
\end{equation}
Analogously conservation of constraints $T_{(\zeta)}\approx0$ determines Lagrange multipliers $\pi$
\begin{equation}
\pi=-{\textstyle\frac{\varrho^{\sigma}}{\varrho^{\tau}}}\partial_\sigma\zeta-\bar k_{\mathcal W}\zeta+{\textstyle\frac{2ic\alpha' s_{\bar{\mathcal Z}\mathcal W}}{\varrho^{\tau}}}\eta,\quad
\bar\pi=-{\textstyle\frac{\varrho^{\sigma}}{\varrho^{\tau}}}\partial_\sigma\bar\zeta-k_{\mathcal W}\bar\zeta-{\textstyle\frac{2ic\alpha' s_{\bar{\mathcal W}\mathcal Z}}{\varrho^{\tau}}}\bar\eta.
\end{equation}
The conservation of the primary constraints $P^{(\rho)}_\tau\approx0$, $P^{(\varrho)}_\tau\approx0$, $\chi_{\mathcal Z}\approx0$, $\chi_{\mathcal W}\approx0$, as well as, the secondary constraints $T_{\mathcal Z\sigma}\approx0$ and $T_{\mathcal W\sigma}\approx0$ does not lead neither to new equations for Lagrange multipliers nor to extra constraints, whereas conservation of the orthogonality relations $\chi_{\bar{\mathcal W}\mathcal Z}\approx0$ and $\chi_{\bar{\mathcal Z}\mathcal W}\approx0$ yields pair of the secondary constraints
\begin{equation}\label{13}
\omega_{\bar{\mathcal W}\mathcal Z}=\bar{\mathcal W}_A\partial_\sigma\mathcal Z^A-\partial_\sigma\bar{\mathcal W}_A\mathcal Z^A\approx0,\quad\omega_{\bar{\mathcal Z}\mathcal W}=\partial_\sigma\bar{\mathcal Z}_A\mathcal W^A-\bar{\mathcal Z}_A\partial_\sigma\mathcal W^A\approx0,
\end{equation}
 which conservation in turn fixes Lagrange multipliers $s_{\bar{\mathcal W}\mathcal Z}$ and $s_{\bar{\mathcal Z}\mathcal W}$
\begin{equation}
s_{\bar{\mathcal W}\mathcal Z}={\textstyle\frac{1}{c\alpha'}}\left({\textstyle\frac{\varrho^{\sigma}}{\varrho^{\tau}}}-{\textstyle\frac{\rho^{\sigma}}{\rho^{\tau}}}\right)F,\quad
s_{\bar{\mathcal Z}\mathcal W}={\textstyle\frac{1}{c\alpha'}}\left({\textstyle\frac{\varrho^{\sigma}}{\varrho^{\tau}}}-{\textstyle\frac{\rho^{\sigma}}{\rho^{\tau}}}\right)\bar F,
\end{equation}
where
\begin{equation}
F={\textstyle\frac14}\left({\textstyle\frac{1}{n\bar n}}\partial_\sigma\bar{\mathcal Z}_A\partial_\sigma\mathcal W^A+{\textstyle\frac{i\rho^{\tau}}{n}}\partial_\sigma u^\alpha u_\alpha+{\textstyle\frac{i\varrho^{\tau}}{\bar n}}\bar v^{\dot\alpha}\partial_\sigma\bar v_{\dot\alpha}\right).
\end{equation}

We thus arrive at the following expression for the total Hamiltonian density as the linear combination of the first-class constraints with the Lagrange multipliers that remained undetermined
\begin{equation}
H_T(\tau,\sigma)={\textstyle\frac{\rho^{\sigma}}{c\alpha'}}T^{(1)}_{\mathcal Z\sigma}+{\textstyle\frac{\varrho^{\sigma}}{c\alpha'}}T^{(1)}_{\mathcal W\sigma}+s_{\mathcal Z}\chi^{(1)}_{\mathcal Z}+s_{\mathcal W}\chi^{(1)}_{\mathcal W}+\beta_{\mathcal Z}\Delta_{\mathcal Z}+\beta_{\mathcal W}\Delta_{\mathcal W}+\beta^{\sigma(\rho)}P^{(\rho)}_\sigma+\beta^{\sigma(\varrho)}P^{(\varrho)}_\sigma\approx0,
\end{equation}
where
\begin{equation}
\begin{array}{rl}
T^{(1)}_{\mathcal Z\sigma}=&T_{\mathcal Z\sigma}+c\alpha'\partial_\sigma P^{(\rho)}_\tau
-\frac{c\alpha'}{\rho^{\tau}}((\partial_\sigma\mu^\alpha-i\varrho^{\tau}\bar nv^\alpha)T_{(\mu)\alpha}+(\partial_\sigma\bar\mu^{\dot\alpha}+i\varrho^{\tau}n\bar v^{\dot\alpha})\bar T_{(\mu)\dot\alpha}\\[0.2cm]
+& \partial_\sigma u_\alpha T^\alpha_{(u)}+\partial_\sigma\bar u_{\dot\alpha}\bar T^{\dot\alpha}_{(u)}+\partial_\sigma\eta T_{(\eta)}+\partial_\sigma\bar\eta\bar T_{(\eta)})+ic\alpha'(n\bar u^{\dot\alpha}\bar T_{(\nu)\dot\alpha}-\bar nu^\alpha T_{(\nu)\alpha})\\[0.2cm]
-&\frac{F}{\rho^{\tau}}(\chi_{\bar{\mathcal W}\mathcal Z}+\frac{2ic\alpha'}{\rho^{\tau}}(\bar\nu^{\dot\alpha}\bar T_{(\mu)\dot\alpha}+v_\alpha T^\alpha_{(u)}+\zeta T_{(\eta)})-\frac{2ic\alpha'}{\varrho^{\tau}}(\mu^\alpha T_{(\nu)\alpha}+\bar u_{\dot\alpha}\bar T^{\dot\alpha}_{(v)}+\bar\eta\bar T_{(\zeta)}))\\[0.2cm]
-&\frac{\bar F}{\rho^{\tau}}(\chi_{\bar{\mathcal Z}\mathcal W}-\frac{2ic\alpha'}{\rho^{\tau}}(\nu^{\alpha}T_{(\mu)\alpha}+\bar v_{\dot\alpha}\bar T^{\dot\alpha}_{(u)}+\bar\zeta\bar T_{(\eta)})+\frac{2ic\alpha'}{\varrho^{\tau}}(\bar\mu^{\dot\alpha}\bar T_{(\nu)\dot\alpha}+u_{\alpha}T^{\alpha}_{(v)}+\eta T_{(\zeta)}))
\approx0,
\end{array}
\end{equation}
\begin{equation}
\begin{array}{rl}
T^{(1)}_{\mathcal W\sigma}=& T_{\mathcal W\sigma}+c\alpha'\partial_\sigma P^{(\varrho)}_\tau
-\frac{c\alpha'}{\varrho^{\tau}}((\partial_\sigma\nu^\alpha-i\rho^{\tau}\bar nu^\alpha)T_{(\nu)\alpha}+(\partial_\sigma\bar\nu^{\dot\alpha}+i\rho^{\tau}n\bar u^{\dot\alpha})\bar T_{(\nu)\dot\alpha}\\[0.2cm]
+& \partial_\sigma v_\alpha T^\alpha_{(v)}+\partial_\sigma\bar v_{\dot\alpha}\bar T^{\dot\alpha}_{(v)}+\partial_\sigma\zeta T_{(\zeta)}+\partial_\sigma\bar\zeta\bar T_{(\zeta)})+ic\alpha'(n\bar v^{\dot\alpha}\bar T_{(\mu)\dot\alpha}-\bar nv^\alpha T_{(\mu)\alpha})\\[0.2cm]
+&\frac{F}{\varrho^{\tau}}(\chi_{\bar{\mathcal W}\mathcal Z}+\frac{2ic\alpha'}{\rho^{\tau}}(\bar\nu^{\dot\alpha}\bar T_{(\mu)\dot\alpha}+v_\alpha T^\alpha_{(u)}+\zeta T_{(\eta)})-\frac{2ic\alpha'}{\varrho^{\tau}}(\mu^\alpha T_{(\nu)\alpha}+\bar u_{\dot\alpha}\bar T^{\dot\alpha}_{(v)}+\bar\eta\bar T_{(\zeta)}))\\[0.2cm]
+&\frac{\bar F}{\varrho^{\tau}}(\chi_{\bar{\mathcal Z}\mathcal W}-\frac{2ic\alpha'}{\rho^{\tau}}(\nu^{\alpha}T_{(\mu)\alpha}+\bar v_{\dot\alpha}\bar T^{\dot\alpha}_{(u)}+\bar\zeta\bar T_{(\eta)})+\frac{2ic\alpha'}{\varrho^{\tau}}(\bar\mu^{\dot\alpha}\bar T_{(\nu)\dot\alpha}+u_{\alpha}T^{\alpha}_{(v)}+\eta T_{(\zeta)}))
\approx0,
\end{array}
\end{equation}
\begin{equation}
\chi^{(1)}_{\mathcal Z}=\chi_{\mathcal Z}-{\textstyle\frac{2ic\alpha'}{\rho^{\tau}}}(\mu^\alpha T_{(\mu)\alpha}-\bar\mu^{\dot\alpha}\bar T_{(\mu)\dot\alpha}-u_\alpha T^\alpha_{(u)}+\bar u_{\dot\alpha}\bar T^{\dot\alpha}_{(u)}-\eta T_{(\eta)}+\bar\eta\bar T_{(\eta)})\approx0,
\end{equation}
\begin{equation}
\chi^{(1)}_{\mathcal W}=\chi_{\mathcal W}-{\textstyle\frac{2ic\alpha'}{\varrho^{\tau}}}(\nu^\alpha T_{(\nu)\alpha}-\bar\nu^{\dot\alpha}\bar T_{(\nu)\dot\alpha}-v_\alpha T^\alpha_{(v)}+\bar v_{\dot\alpha}\bar T^{\dot\alpha}_{(v)}-\zeta T_{(\zeta)}+\bar\zeta\bar T_{(\zeta)})\approx0,
\end{equation}
\begin{equation}
\Delta_{\mathcal Z}=\mu^\alpha T_{(\mu)\alpha}+\bar\mu^{\dot\alpha}\bar T_{(\mu)\dot\alpha}+u_\alpha T^\alpha_{(u)}+\bar u_\alpha\bar T^{\dot\alpha}_{(u)}+\eta T_{(\eta)}+\bar\eta\bar T_{(\eta)}-2\rho^{\mu}P^{(\rho)}_\mu\approx0,
\end{equation}
\begin{equation}
\Delta_{\mathcal W}=\nu^\alpha T_{\nu\alpha}+\bar\nu^{\dot\alpha}\bar T_{(\nu)\dot\alpha}+v_\alpha T^\alpha_{(v)}+\bar v_\alpha\bar T^{\dot\alpha}_{(v)}+\zeta T_{(\zeta)}+\bar\zeta\bar T_{(\zeta)}-2\varrho^{\mu}P^{(\varrho)}_\mu\approx0
\end{equation}
and the redefinition of Lagrange multipliers was performed $\beta_{\mathcal Z}=-\frac{\beta^{\tau(\rho)}}{2\rho^{\tau}}$, $\beta_{\mathcal W}=-\frac{\beta^{\tau(\varrho)}}{2\varrho^{\tau}}$.

On the Poisson brackets these first-class constraints\footnote{The
above obtained results are also applicable to the case of the
$\kappa-$symmetry gauge fixed $N=1$ closed superstring by putting
to zero either $\zeta, \bar\zeta$ and conjugate momenta, or $\eta,
\bar\eta$ and their momenta depending on the sign $s$ of the
Wess-Zumino term of the original action. For the bosonic string it
is necessary to put to zero both $\eta, \bar\eta$ and $\zeta,
\bar\zeta$ with the conjugate momenta.} generate gauge
symmetries of the superstring action (\ref{4}). $T^{(1)}_{\mathcal
Z\sigma}\approx0$ and $T^{(1)}_{\mathcal W\sigma}\approx0$
constraints are Virasoro generators and correspond to the
world-sheet reparametrizations. $\Delta_{\mathcal
Z}\approx0$ constraint generates dilatation of the $\mathcal Z^A$
supertwistor components
\begin{equation}
\begin{array}{c}
\delta_{d_{\mathcal Z}}\mathcal Z^A=d_{\mathcal Z}\mathcal Z^A,\quad \delta_{d_{\mathcal Z}}\bar{\mathcal Z}_A=d_{\mathcal Z}\bar{\mathcal Z}_A,\quad\delta_{d_{\mathcal Z}}\rho^{\mu}=-2d_{\mathcal Z}\rho^{\mu},
\end{array}
\end{equation}
whereas the $\Delta_{\mathcal W}\approx0$ constraint generates independent dilatation of the $\mathcal W^A$ supertwistor components
\begin{equation}
\begin{array}{c}
\delta_{d_{\mathcal W}}\mathcal W^A=d_{\mathcal W}\mathcal W^A,\quad \delta_{d_{\mathcal W}}\bar{\mathcal W}_A=d_{\mathcal W}\bar{\mathcal W}_A,\quad
\delta_{d_{\mathcal W}}\varrho^{\mu}=-2d_{\mathcal W}\varrho^{\mu}.
\end{array}
\end{equation}
The fact that these symmetries are independent ones is due to the utilization of the unnormalized dyad. They can be treated as the combination of Weyl and local $SO(1,1)$ transformations if one identifies $d_{\mathcal Z}=\varsigma-\varrho$, $d_{\mathcal W}=\varsigma+\varrho$. The $\chi^{(1)}_{\mathcal Z}\approx0$ constraint generates local $U(1)_{\mathcal Z}$ rotation of the $\mathcal Z^A$ supertwistor components
\begin{equation}
\delta_{\varphi_{\mathcal Z}}\mathcal Z^A=i\varphi_{\mathcal Z}\mathcal Z^A,\quad \delta_{\varphi_{\mathcal Z}}\bar{\mathcal Z}_A=-i\varphi_{\mathcal Z}\bar{\mathcal Z}_A.
\end{equation}
Similarly the $\chi^{(1)}_{\mathcal W}\approx0$ constraint generates local $U(1)_{\mathcal W}$ rotation of the $\mathcal W^A$ supertwistor components
\begin{equation}
\delta_{\varphi_{\mathcal W}}\mathcal W^A=i\varphi_{\mathcal W}\mathcal W^A,\quad \delta_{\varphi_{\mathcal W}}\bar{\mathcal W}_A=-i\varphi_{\mathcal W}\bar{\mathcal W}_A.
\end{equation}

All the other constraints are the second-class ones. These are constraints (\ref{10}), (\ref{11}) and (\ref{7}),
(\ref{13}) that should be taken into account by constructing the Dirac brackets (D.B.). This is easily accomplished for the primary constraints (\ref{10}), (\ref{11}), whose algebra is characterized by the following nonzero P.B. relations
\begin{equation}
\begin{array}{c}
\{T_{(\mu)\alpha}(\sigma),T^\beta_{(u)}(\sigma')\}_{P.B.}=\frac{i\rho^{\tau}}{2c\alpha'}\delta_\alpha^\beta\delta(\sigma-\sigma'),\
\{\bar T_{(\mu)\dot\alpha}(\sigma),\bar T^{\dot\beta}_{(u)}(\sigma')\}_{P.B.}=-\frac{i\rho^{\tau}}{2c\alpha'}\delta_{\dot\alpha}^{\dot\beta}\delta(\sigma-\sigma'),\\[0.2cm]
\{T_{(\nu)\alpha}(\sigma),T^\beta_{(v)}(\sigma')\}_{P.B.}=\frac{i\varrho^{\tau}}{2c\alpha'}\delta_\alpha^\beta\delta(\sigma-\sigma'),\
\{\bar T_{(\nu)\dot\alpha}(\sigma),\bar T^{\dot\beta}_{(v)}(\sigma')\}_{P.B.}=-\frac{i\varrho^{\tau}}{2c\alpha'}\delta_{\dot\alpha}^{\dot\beta}\delta(\sigma-\sigma'),\\[0.2cm]
\{T_{(\eta)}(\sigma),\bar T_{(\eta)}(\sigma')\}_{P.B.}=\frac{i\rho^{\tau}}{2c\alpha'}\delta(\sigma-\sigma'),\quad
\{T_{(\zeta)}(\sigma),\bar T_{(\zeta)}(\sigma')\}_{P.B.}=\frac{i\varrho^{\tau}}{2c\alpha'}\delta(\sigma-\sigma').
\end{array}
\end{equation}
So that the Dirac matrix ${\mathbf C}_{\mathcal{FG}}$ has the following block-diagonal structure
\begin{equation}
{\mathbf C}_{\mathcal{FG}}\sim\begin{array} {l|cccccccccccc|}
 \multicolumn{1}{c}{}
& \multicolumn{1}{c}{T_{(\mu)\beta}}&
T^\beta_{(u)}&
\bar T_{(\mu)\dot\beta} &
\bar T^{\dot\beta}_{(u)} &
T_{(\nu)\beta} &
T^\beta_{(v)} &
\bar T_{(\nu)\dot\beta} &
\bar T^{\dot\beta}_{(v)} &
T_{(\eta)} &
\bar T_{(\eta)} &
T_{(\zeta)} &
\multicolumn{1}{c}{\bar T_{(\zeta)}}\\[0.2cm] \cline{2-13}
T_{(\mu)\alpha} &0&\multicolumn{1}{c|}{\delta_\alpha^\beta}&&&&&&&&&&\\[0.2cm]
T^\alpha_{(u)}&-\delta^\alpha_\beta &\multicolumn{1}{c|}{0}&&&&&&&&&&\\[0.2cm] \cline{2-5}
\bar T_{(\mu)\dot\alpha}&&&\multicolumn{1}{|c}{0}&\multicolumn{1}{c|}{-\delta_{\dot\alpha}{}^{\dot\beta}}&&&&&&&&\\[0.2cm]
\bar T^{\dot\alpha}_{(u)} &&&\multicolumn{1}{|c}{\delta^{\dot\alpha}_{\dot\beta}}&\multicolumn{1}{c|}{0}&&&&&{\mathbf 0}&&&\\[0.2cm] \cline{4-7}
T_{(\nu)\alpha} &&&&&\multicolumn{1}{|c}{0}&\multicolumn{1}{c|}{\delta_\alpha^\beta}&&&&&&\\[0.2cm]
T^\alpha_{(v)} &&&&&\multicolumn{1}{|c}{-\delta^\alpha_\beta}&\multicolumn{1}{c|}{0}&&&&&&\\[0.2cm] \cline{6-9}
\bar T_{(\nu)\dot\alpha} &&&&&&&\multicolumn{1}{|c}{0}&\multicolumn{1}{c|}{-\delta_{\dot\alpha}^{\dot\beta}}&&&&\\[0.2cm]
\bar T^{\dot\alpha}_{(v)}  &&&{\mathbf 0}&&&&\multicolumn{1}{|c}{\delta^{\dot\alpha}_{\dot\beta}}&\multicolumn{1}{c|}{0}&&&&\\[0.2cm]
\cline{8-11}
T_{(\eta)} &&&&&&&&&\multicolumn{1}{|c}{0}&\multicolumn{1}{c|}{1}&&\\[0.2cm]
\bar T_{(\eta)} &&&&&&&&&\multicolumn{1}{|c}{1}&\multicolumn{1}{c|}{0}&& \\ \cline{10-13}
T_{(\zeta)} &&&&&&&&&&&\multicolumn{1}{|c}{0}&\multicolumn{1}{c|}{1}\\[0.2cm]
\bar T_{(\zeta)} &&&&&&&&&&&\multicolumn{1}{|c}{1}&\multicolumn{1}{c|}{0}\\
\cline{2-13}
\end{array}\ .
\end{equation}
The inverse Dirac matrix enters the D.B. definition for a pair of phase-space functions
\begin{equation}
\begin{array}{rl}
\{f(\sigma),g(\sigma^\prime)\}_{D.B.}=&\{f(\sigma ),g(\sigma^\prime)\}_{P.B.} \\[0.2cm]
-&
\int d\sigma''d\sigma'''
\{f(\sigma ),\mathcal F(\sigma'')\}_{P.B.}
( {\mathbf C}^{-1})^{\mathcal{FG}}(\sigma'',\sigma''')\{\mathcal G(\sigma''')\,,
g(\sigma^\prime)\}_{P.B.},
\end{array}
\end{equation}
where $\mathcal F$ and $\mathcal G$ collectively denote the second-class constraints forming ${\mathbf C}_{\mathcal{FG}}$.
This definition results in the following nonzero D.B. for the supertwistor components
\begin{equation}\label{12}
\{\mathcal Z^A(\sigma),\bar{\mathcal Z}_B(\sigma')\}_{D.B.}={\textstyle\frac{2ic\alpha'}{\rho^{\tau}}}\delta^A_B\delta(\sigma-\sigma'),\quad
\{\mathcal W^A(\sigma),\bar{\mathcal W}_B(\sigma')\}_{D.B.}={\textstyle\frac{2ic\alpha'}{\varrho^{\tau}}}\delta^A_B\delta(\sigma-\sigma').
\end{equation}
The presence of the $\tau$-components of the world-sheet density ($\rho{}^\mu$, $\varrho{}^\mu$) in the second-class constraints (\ref{10}), (\ref{11}) results in the nonzero D.B. between $P^{(\rho,\varrho)}_\tau$ and the supertwistor components
\begin{equation}
\begin{array}{c}
\{\mathcal Z^A(\sigma),P^{(\rho)}_\tau(\sigma')\}_{D.B.}=\frac{1}{2\rho^{\tau}}\mathcal Z^A\delta(\sigma-\sigma'),\
\{\bar{\mathcal Z}_A(\sigma),P^{(\rho)}_\tau(\sigma')\}_{D.B.}=\frac{1}{2\rho^{\tau}}\bar{\mathcal Z}_A\delta(\sigma-\sigma'),\\[0.2cm]
\{\mathcal W^A(\sigma),P^{(\varrho)}_\tau(\sigma')\}_{D.B.}=\frac{1}{2\varrho^{\tau}}\mathcal W^A\delta(\sigma-\sigma'),\
\{\bar{\mathcal W}_A(\sigma),P^{(\varrho)}_\tau(\sigma')\}_{D.B.}=\frac{1}{2\varrho^{\tau}}\bar{\mathcal W}_A\delta(\sigma-\sigma').\\[0.2cm]
\end{array}
\end{equation}
Then remaining second-class constraints (\ref{7}), (\ref{13}) satisfy the following nonzero D.B. relations
\begin{equation}
\begin{array}{c}
\{\chi_{\bar{\mathcal W}\mathcal Z}(\sigma),\chi_{\bar{\mathcal Z}\mathcal W}(\sigma')\}_{D.B.}=2ic\alpha'\left(\frac{\chi_{\mathcal W}}{\rho^{\tau}}-\frac{\chi_{\mathcal Z}}{\varrho^{\tau}}\right)\delta(\sigma-\sigma'),\\[0.2cm]
\{\chi_{\bar{\mathcal Z}\mathcal W}(\sigma),\omega_{\bar{\mathcal W}\mathcal Z}(\sigma')\}_{D.B.}=2ic\alpha'\left(\frac{\chi_{\mathcal Z}(\sigma')}{\varrho^{\tau}(\sigma)}+\frac{\chi_{\mathcal W}(\sigma')}{\rho^{\tau}(\sigma)}\right)\partial_\sigma\delta(\sigma-\sigma')-\mathcal T\delta(\sigma-\sigma'),
\\[0.2cm]
\{\chi_{\bar{\mathcal W}\mathcal Z}(\sigma),\omega_{\bar{\mathcal Z}\mathcal W}(\sigma')\}_{D.B.}=-2ic\alpha'\left(\frac{\chi_{\mathcal Z}(\sigma')}{\varrho^{\tau}(\sigma)}+\frac{\chi_{\mathcal W}(\sigma')}{\rho^{\tau}(\sigma)}\right)\partial_\sigma\delta(\sigma-\sigma')-\mathcal T\delta(\sigma-\sigma'),\\[0.2cm]
\{\omega_{\bar{\mathcal W}\mathcal Z}(\sigma),\omega_{\bar{\mathcal Z}\mathcal W}(\sigma')\}_{D.B.}=2ic\alpha'\left(\frac{\chi_{\mathcal Z}}{\varrho^{\tau}}-\frac{\chi_{\mathcal W}}{\rho^{\tau}}\right)(\sigma)\partial^2_\sigma\delta(\sigma-\sigma')\\[0.2cm]
+2ic\alpha'\partial_\sigma\left(\frac{\chi_{\mathcal Z}}{\varrho^{\tau}}-\frac{\chi_{\mathcal W}}{\rho^{\tau}}\right)(\sigma)\partial_\sigma\delta(\sigma-\sigma')
+2ic\alpha'\partial_\sigma\left(\frac{\partial_\sigma\chi_{\mathcal Z}}{\varrho^{\tau}}-\frac{\partial_\sigma\chi_{\mathcal W}}{\rho^{\tau}}\right)\delta(\sigma-\sigma')\\[0.2cm]
+\mathcal T(\sigma)\partial_\sigma\delta(\sigma-\sigma')-\mathcal T(\sigma')\partial_{\sigma'}\delta(\delta-\delta')
+8ic\alpha'\left(\frac{1}{\rho^{\tau}}\partial_\sigma\bar{\mathcal W}_A\partial_\sigma\mathcal W^A-\frac{1}{\varrho^{\tau}}\partial_\sigma\bar{\mathcal Z}_A\partial_\sigma\mathcal Z^A\right)\delta(\sigma-\sigma'),
\end{array}
\end{equation}
where $\mathcal T=8c\alpha'(\frac{T_{\mathcal Z\sigma}}{\varrho^{\tau}}-\frac{T_{\mathcal W\sigma}}{\rho^{\tau}}-n\bar n)$.

Upon considering the constraints (\ref{10}), (\ref{11}) as strong equalities the form of the first-class constraints  reduces to (\ref{10'}), (\ref{11'}), (\ref{6}) and
\begin{equation}
\tilde T_{\mathcal Z\sigma}=T_{\mathcal Z\sigma}-{\textstyle\frac{F}{\rho^{\tau}}}\chi_{\bar{\mathcal W}\mathcal Z}-{\textstyle\frac{\bar F}{\rho^{\tau}}}\chi_{\bar{\mathcal Z}\mathcal W}\approx0,\quad
\tilde T_{\mathcal W\sigma}=T_{\mathcal W\sigma}+{\textstyle\frac{F}{\varrho^{\tau}}}\chi_{\bar{\mathcal W}\mathcal Z}+{\textstyle\frac{\bar F}{\varrho^{\tau}}}\chi_{\bar{\mathcal Z}\mathcal W}\approx0.
\end{equation}
Using the above introduced D.B. one can calculate the D.B. algebra of the first-class constraints
\begin{equation}
\begin{array}{c}
\{P^{(\rho)}_\tau(\sigma),\chi_{\mathcal Z}(\sigma')\}_{D.B.}=-\frac{1}{\rho^{\tau}}\chi_{\mathcal Z}\delta(\sigma-\sigma'),\\[0.2cm]
\{P^{(\rho)}_\tau(\sigma),\tilde T_{\mathcal Z\sigma}(\sigma')\}_{D.B.}=-\frac{1}{\rho^{\tau}}\tilde T_{\mathcal Z\sigma}\delta(\sigma-\sigma')+O(\chi_{\bar{\mathcal W}\mathcal Z}, \chi_{\bar{\mathcal Z}\mathcal W}),\\[0.2cm]
\{\chi_{\mathcal Z}(\sigma),\tilde T_{\mathcal Z\sigma}(\sigma')\}_{D.B.}=\frac{c\alpha'}{\rho^{\tau}(\sigma)}\chi_{\mathcal Z}(\sigma')\partial_\sigma\delta(\sigma-\sigma')+O(\chi_{\bar{\mathcal W}\mathcal Z}, \chi_{\bar{\mathcal Z}\mathcal W}),\\[0.2cm]
\{\tilde T_{\mathcal Z\sigma}(\sigma),\tilde T_{\mathcal Z\sigma}(\sigma')\}_{D.B.}=\frac{c\alpha'}{\rho^{\tau}(\sigma)}\tilde T_{\mathcal Z\sigma}(\sigma)\partial_\sigma\delta(\sigma-\sigma')-
\frac{c\alpha'}{\rho^{\tau}(\sigma')}\tilde T_{\mathcal Z\sigma}(\sigma')\partial_{\sigma'}\delta(\sigma-\sigma')
+O(\chi, \chi^2).
\end{array}
\end{equation}
Analogous relations hold for the constraints from the $\mathcal W$-sector.

Introducing D.B. that take into account the second-class constraints (\ref{7}), (\ref{13}) makes the first-class constraint algebra even more complicated. Although partial simplification can be achieved by using the reparametrization, local $SO(1,1)$ and Weyl symmetries of the action (\ref{4}) to gauge fix the world-sheet density components, for instance, as
\begin{equation}\label{14}
\rho^{\tau}=\varrho^\tau=1,\ \rho^{\sigma}=-\varrho^\sigma=1.
\end{equation}
The reason for rather complicated gauge symmetry algebra of the
superstring in supertwistor formulation is the presence of the
constraints (\ref{7}), (\ref{13}) that mix supertwistor variables
from the $\mathcal Z$- and $\mathcal W$-sectors and are necessary
to establish the correspondence with the superspace-time
description. Alternatively one can try to take into account these
second-class constraints by converting them into some effective
first-class ones in the phase-space enlarged by conversion
variables. We plan to examine such a possibility but here, as the
preliminary step, we just relax these constraints and consider the
resulting model. Such modification amounts to considering the
superstring model in the complexified superspace \cite{Shirafuji} or by introducing
two independent sets of superspace coordinates each of which is
associated with either $\mathcal Z$ or $\mathcal W$ supertwistors \cite{Berkovits}, \cite{BerkovitsMotl}.

Then the D.B. algebra of the first-class constraints, that now coincide with the primary (\ref{6}) and secondary (\ref{8}), (\ref{9}) ones, reduces to the following nonzero relations
\begin{equation}
\begin{array}{c}
\{T_{\mathcal Z(\mathcal W)\sigma}(\sigma),\chi_{\mathcal Z(\mathcal W)}(\sigma')\}_{D.B.}=c\alpha'\chi_{\mathcal Z(\mathcal W)}(\sigma)\partial_\sigma\delta(\sigma-\sigma'),\\[0.2cm]
\{T_{\mathcal Z(\mathcal W)\sigma}(\sigma),T_{\mathcal Z(\mathcal W)\sigma}(\sigma')\}_{D.B.}=c\alpha'T_{\mathcal Z(\mathcal W)\sigma}(\sigma)\partial_\sigma\delta(\sigma-\sigma')-c\alpha'T_{\mathcal Z(\mathcal W)\sigma}(\sigma')\partial_{\sigma'}\delta(\sigma-\sigma').
\end{array}
\end{equation}
Passing from the D.B.'s to quantum equal-time (anti)commutators
and treating the Virasoro and $U(1)$ symmetry generators as
operators, e.g. creation-annihilation ordered, it is possible to
evaluate quantum symmetry algebra of the superstring. On the other
hand, since the superstring action based on the Lagrangian
(\ref{4}) is Weyl invariant these results can be obtained by
applying the world-sheet CFT technique that is the subject of the
next section. It is also preferable to use world-sheet CFT from the perspective of comparison to existing twistor superstring models.

\section{Quantum symmetries of $D=4$ superstrings in supertwistor formulation and generalizations}

Upon gauge-fixing reparametrization, local $SO(1,1)$ and Weyl symmetries and relaxing the second-class constraints (\ref{7}), (\ref{13}) the action (\ref{4}) reduces to
\begin{equation}\label{15}
S=\!-{\textstyle\frac{i}{2c\alpha'}}\!\!\int\!\! d^2\xi(\partial_{+2}\bar{\mathcal Z}_A\mathcal Z^A-\bar{\mathcal Z}_A\partial_{+2}\mathcal Z^A+\partial_{-2}\bar{\mathcal W}_A\mathcal W^A-\bar{\mathcal W}_A\partial_{-2}\mathcal W^A-2in\bar n)+\!\int\!\! d^2\xi(s_{\mathcal Z}\chi_{\mathcal Z}+s_{\mathcal W}\chi_{\mathcal W})
\end{equation}
and should be supplemented by the Virasoro generators
\begin{equation}
T_{+2+2}\!=\! i(\partial_{-2}\bar{\mathcal Z}_A\mathcal Z^A-\bar{\mathcal Z}_A\partial_{-2}\mathcal Z^A)+4n\bar n=0,\quad
T_{-2-2}\!=\! i(\partial_{+2}\bar{\mathcal W}_A\mathcal W^A-\bar{\mathcal W}_A\partial_{+2}\mathcal W^A)+4n\bar n=0.
\end{equation}
This is not the action of a free $2d$ theory but rather contains
the interaction term $n\bar
n=u^\alpha\varepsilon_{\alpha\beta}v^\beta\bar
u^{\dot\alpha}\varepsilon_{\dot\alpha\dot\beta}\bar v^{\dot\beta}$
intertwining the components of different supertwistors. It is this term that
reduces global $D=4$ superconformal symmetry of the action down to
superPoincare one because of the presence of
$\varepsilon_{\alpha\beta}$, $\varepsilon_{\dot\alpha\dot\beta}$
tensors that are not conformally covariant quantities. This
interaction term can be identified with the mass squared operator introduced in twistor theory \cite{twistor}.
Indeed a light-like 4-momentum $p_{\alpha\dot\alpha}$
can be presented as the product of $SL(2,\mathbb C)$ spinor and its conjugate
$p^{(u)}_{\alpha\dot\alpha}=\frac{1}{(\alpha')^{1/2}}u_\alpha\bar
u_{\dot\alpha}$ or
$p^{(v)}_{\alpha\dot\alpha}=\frac{1}{(\alpha')^{1/2}}v_\alpha\bar
v_{\dot\alpha}$. 4-Momentum of a massive state can be given modulo overall normalization by the
sum
$p_{\alpha\dot\alpha}=p^{(u)}_{\alpha\dot\alpha}+p^{(v)}_{\alpha\dot\alpha}$,
so that $p^2=\frac{n\bar n}{\alpha'}$. So the interaction term
in the superstring action (\ref{15}) is the twistor representation of the
conventional string momentum squared term
$p^2=\frac{2}{\alpha'}\alpha^2_0=\frac{2}{\alpha'}\tilde\alpha^2_0$
that is the zero mode contribution to the Virasoro generators
$L_0$, $\bar L_0$ and 'measures' the mass of string states. In the
space-time formulation this mass term arises from the free
field $\partial X^m\partial X_m$ contribution to the energy-momentum
tensor but in the twistor description, because of the Penrose
representation for the 4-momenta, it appears to be of the
fourth order in the supertwistor components. Hence to apply the world-sheet CFT technique we have to linearize the action (\ref{15}).

Passing to Euclidean world sheet the kinetic term of the linearized superstring action can be brought to the form
\begin{equation}\label{16}
S=-\frac{i}{2c\alpha'}\int dzd\bar z(\bar\partial\bar{\cal Z}_A{\cal Z}^A-\bar{\cal Z}_A\bar\partial{\cal Z}^A+\partial\bar{\mathcal W}_A{\mathcal W}^A-\bar{\mathcal W}_A\partial{\mathcal W}^A).
\end{equation}
So that the supertwistor components are characterized by the free-field OPEs
\begin{equation}
\mathcal Z^A(z_1)\bar{\mathcal Z}_B(z_2)\sim\frac{ic\alpha'}{z_{12}}\delta^A_B,\quad\mathcal W^A(\bar z_1)\bar{\mathcal W}_B(\bar z_2)\sim\frac{ic\alpha'}{\bar z_{12}}\delta^A_B.
\end{equation}
Holomorphic Virasoro generator
\begin{equation}
T_{\mathcal Z}(z)=-\frac{i}{2c\alpha'}(\partial\bar{\cal Z}_A{\cal Z}^A-\bar{\cal Z}_A\partial{\cal Z}^A)
\end{equation}
has the following OPEs with itself and with the $U(1)_{\mathcal Z}$ symmetry generator $\chi_{\mathcal Z}(z)=\frac{1}{c\alpha'}\bar{\mathcal Z}_A\mathcal Z^A$
\begin{equation}\label{18}
\begin{array}{c}
\chi_{\mathcal Z}(z_1)\chi_{\mathcal Z}(z_2)\sim\frac{(-4+1)}{z^2_{12}},\quad T_{\mathcal Z}(z_1)\chi_{\mathcal Z}(z_2)\sim\frac{\chi_{\mathcal Z}(z_2)}{z^2_{12}}+\frac{\partial\chi_{\mathcal Z}}{z_{12}},\\[0.2cm]
T_{\mathcal Z}(z_1)T_{\mathcal Z}(z_2)\sim\frac{(-4+1)}{2z^4_{12}}+\frac{2T_{\mathcal Z}(z_2)}{z^2_{12}}+\frac{\partial T_{\mathcal Z}}{z_{12}}.
\end{array}
\end{equation}
In the above expressions anomalous contribution of bosonic components of the supertwistor, equal to -4, is given separately from the fermionic component contribution, equal to 1. Operators $T_{\mathcal Z}(z)$ and $\chi_{\mathcal Z}(z)$ can be considered as bosonic generators of the holomorphic $N=2$ superconformal algebra.
The form of the Virasoro generator $T_{\mathcal Z}(z)$ implies that the $\mathcal Z^A$ supertwistor and its dual $\bar{\mathcal Z}_A$ have  conformal weight $(\frac12,0)$
\begin{equation}
T_{\mathcal Z}(z_1){\cal Z}^A(z_2)\sim\frac{1}{2}\frac{{\cal Z}^A(z_2)}{z^2_{12}}+\frac{\partial{\cal Z}^A}{z_{12}},\quad T_{\mathcal Z}(z_1)\bar{\cal Z}_A(z_2)\sim\frac{1}{2}\frac{\bar{\cal Z}_A(z_2)}{z^2_{12}}+\frac{\partial\bar{\cal Z}_A}{z_{12}}.
\end{equation}

In the antiholomorphic sector $U(1)_{\mathcal W}$ symmetry
generator $\chi_{\mathcal W}(\bar
z)=\frac{1}{c\alpha'}\bar{\mathcal W}_A\mathcal W^A$ and antiholomorphic Virasoro generator
\begin{equation}
T_{\mathcal W}(\bar z)=-\frac{i}{2c\alpha'}(\bar\partial\bar{\mathcal W}_A{\mathcal W}^A-\bar{\mathcal W}_A\bar\partial{\mathcal W}^A)
\end{equation}
satisfy by the following OPEs
\begin{equation}\label{19}
\begin{array}{c}
\chi_{\mathcal W}(\bar z_1)\chi_{\mathcal W}(\bar z_2)\sim\frac{(-4+1)}{\bar z^2_{12}},\quad T_{\mathcal W}(\bar z_1)\chi_{\mathcal W}(\bar z_2)\sim\frac{\chi_{\mathcal W}(\bar z_2)}{\bar z^2_{12}}+\frac{\bar\partial\chi_{\mathcal W}}{\bar z_{12}},\\[0.3cm]
T_{\mathcal W}(\bar z_1)T_{\mathcal W}(\bar z_2)\sim\frac{(-4+1)}{2\bar z^4_{12}}+\frac{2T_{\mathcal W}(\bar z_2)}{\bar z^2_{12}}+\frac{\bar\partial T_{\mathcal W}}{\bar z_{12}}.
\end{array}
\end{equation}
So that the supertwistor $\mathcal W^A$ and its dual $\bar{\mathcal W}_A$ have conformal weight $(0,\frac12)$
\begin{equation}
T_{\mathcal W}(\bar z_1){\mathcal W}^A(\bar z_2)\sim\frac{1}{2}\frac{{\mathcal W}^A(\bar z_2)}{\bar z^2_{12}}+\frac{\bar\partial{\mathcal W}^A}{\bar z_{12}},\quad T_{\mathcal W}(\bar z_1)\bar{\mathcal W}_A(\bar z_2)\sim\frac{1}{2}\frac{\bar{\mathcal W}_A(\bar z_2)}{\bar z^2_{12}}+\frac{\bar\partial\bar{\mathcal W}_A}{\bar z_{12}}.
\end{equation}

The above results can be generalized to the case of $N$-extended supersymmetry, so that the supertwistors are defined as
\begin{equation}
\mathfrak Z^{\mathcal A}=(\mu^\alpha, \bar u_{\dot\alpha}, \bar\eta^i),\quad\bar{\mathfrak Z}_{\mathcal A}=(u_\alpha, \bar\mu^{\dot\alpha}, \eta_i),\quad i,j=1,...,N
\end{equation}
and
\begin{equation}
\mathfrak W^{\mathcal A}=(\nu^\alpha, \bar v_{\dot\alpha}, \bar\zeta^{i'}),\quad\bar{\mathfrak W}_{\mathcal A}=(v_\alpha, \bar\nu^{\dot\alpha}, \zeta_{i'}),\quad i',j'=1,...,N'.
\end{equation}
For the closed superstring $N'$ is not necessarily equal to $N$, while for the open superstring holomorphic and antiholomorphic components can be identified at the world-sheet boundary $\partial\rm M^2$: $\mathfrak Z^{\mathcal A}|_{\partial\rm M^2}=\mathfrak W^{\mathcal A}|_{\partial\rm M^2}$, $\bar{\mathfrak Z}_{\mathcal A}|_{\partial\rm M^2}=\bar{\mathfrak W}_{\mathcal A}|_{\partial\rm M^2}$ implying that $N=N'$.

Holomorphic sector $U(1)_{\mathfrak Z}$ generator
\begin{equation}
\chi_{\mathfrak Z}(z)=\frac{1}{c\alpha'}\bar{\mathfrak Z}_{\mathcal A}{\mathfrak Z}^{\mathcal A}
\end{equation}
and Virasoro generator
\begin{equation}
T_{\mathfrak Z}(z)=-\frac{i}{2c\alpha'}(\bar{\mathfrak Z}_{\mathcal A}\partial{\mathfrak Z}^{\mathcal A}-\partial\bar{\mathfrak Z}_{\mathcal A}{\mathfrak Z}^{\mathcal A})
\end{equation}
satisfy now the following OPEs
\begin{equation}\label{20}
\begin{array}{c}
\chi_{\mathfrak Z}(z_1)\chi_{\mathfrak Z}(z_2)\sim\frac{(N-4)}{z^2_{12}},\quad T_{\mathfrak Z}(z_1)\chi_{\mathfrak Z}(z_2)\sim\frac{\chi_{\mathfrak Z}(z_2)}{z^2_{12}}+\frac{\partial\chi_{\mathfrak Z}}{z_{12}},\\[0.3cm]
T_{\mathfrak Z}(z_1)T_{\mathfrak Z}(z_2)\sim\frac{(N-4)}{2z^4_{12}}+\frac{2T_{\mathfrak Z}(z_2)}{z^2_{12}}+\frac{\partial T_{\mathfrak Z}}{z_{12}}
\end{array}
\end{equation}
with the central charge $C=N-4$. Corresponding expressions hold for the antiholomorphic generators with the conformal anomaly value $\tilde C=N'-4$.

It is seen that the $N=4$ superstring is special because of cancellation between the contributions of bosonic and fermionic components so that the supertwistor variables do not contribute to conformal and $U(1)$ anomalies. In general one has to add extra matter contributing $4-N$ to the $U(1)$ anomaly and $32-N$ to the conformal anomaly in order to compensate contributions of the reparametrization and U(1) ghosts. Although the string models based on the supertwistor variables with $N>1$ can not be obtained from the Green-Schwarz superstring in $D=4$, $1<N\leq4$ case can be related to the reductions of higher dimensional superstrings (see e.g. \cite{Zunger}, \cite{BdAM}).

Above we considered the case of the conjugate supertwistors on each side of the string, whose conformal weights equal $(\frac12,0)$ and $(0,\frac12)$. It could be of interest from the perspective of a possible comparison with the nontopological twistor superstrings \cite{Berkovits}, \cite{BerkovitsMotl} to generalize the above models to the case when the whole $GL(1,\mathbb C)$ is gauged rather than its $U(1)$ subgroup. To this end we introduce the second pair of supertwistors with the following action that generalizes (\ref{16})
\begin{equation}\label{21}
S=-\frac{i}{2c\alpha'}\int dzd\bar z(\bar\partial\bar{\mathfrak Z}_{1\mathcal A}{\mathfrak Z}_2^{\mathcal A}-\bar{\mathfrak Z}_{2\mathcal A}\bar\partial{\mathfrak Z}_1^{\mathcal A}+
\partial\bar{\mathfrak W}_{1\mathcal A}{\mathfrak W}_2^{\mathcal A}-\bar{\mathfrak W}_{2\mathcal A}\partial{\mathfrak W}_1^{\mathcal A}
+(1\leftrightarrow2)),
\end{equation}
where $\bar{\mathfrak Z}_{1\mathcal A}$ is dual to $\mathfrak Z^{\mathcal A}_1$, $\bar{\mathfrak Z}_{2\mathcal A}$ to $\mathfrak Z^{\mathcal A}_2$ and respectively for the $\mathfrak W$-supertwistors.
In the holomorphic sector corresponding Virasoro generator equals
\begin{equation}
T_{\mathfrak Z}(z)=-\frac{i}{2c\alpha'}(\partial\bar{\mathfrak Z}_{1\mathcal A}{\mathfrak Z}_2^{\mathcal A}-\bar{\mathfrak Z}_{2\mathcal A}\partial{\mathfrak Z}_1^{\mathcal A}+(1\leftrightarrow2))
\end{equation}
and has to be supplemented by two extra currents
\begin{equation}
\psi_{\mathfrak Z}(z)=\frac{1}{c\alpha'}(\bar{\mathfrak Z}_{1\mathcal A}\mathfrak Z^{\mathcal A}_2+\bar{\mathfrak Z}_{2\mathcal A}\mathfrak Z^{\mathcal A}_1),\quad
\chi_{\mathfrak Z}(z)=\frac{i}{c\alpha'}(\bar{\mathfrak Z}_{1\mathcal A}\mathfrak Z^{\mathcal A}_2-\bar{\mathfrak Z}_{2\mathcal A}\mathfrak Z^{\mathcal A}_1)
\end{equation}
that are required for the action (\ref{21}) to be invariant under the $GL(1,\mathbb C)_{\mathfrak Z}$ gauge symmetry. $\psi_{\mathfrak Z}(z)$ is responsible for the $GL(1,\mathbb R)_{\mathfrak Z}$ part of $GL(1,\mathbb C)_{\mathfrak Z}$ invariance and $\chi_{\mathfrak Z}(z)$ for the $U(1)_{\mathfrak Z}$ part. As a result $\mathfrak Z_{1,2}^{\mathcal A}$ supertwistor components can be viewed as homogeneous coordinates of the projective supertwistor space $\mathbb{CP}^{(3|N)}$. Corresponding generators of the antiholomorphic sector read
\begin{equation}
\begin{array}{c}
T_{\mathfrak W}(\bar z)=-\frac{i}{2c\alpha'}(\bar\partial\bar{\mathfrak W}_{1\mathcal A}{\mathfrak W}_2^{\mathcal A}-\bar{\mathfrak W}_{2\mathcal A}\bar\partial{\mathfrak W}_1^{\mathcal A}+(1\leftrightarrow2)),\\[0.2cm]
\psi_{\mathfrak W}(\bar z)=\frac{1}{c\alpha'}(\bar{\mathfrak W}_{1\mathcal A}\mathfrak W^{\mathcal A}_2+\bar{\mathfrak W}_{2\mathcal A}\mathfrak W^{\mathcal A}_1),\quad\chi_{\mathfrak W}(\bar z)=\frac{i}{c\alpha'}(\bar{\mathfrak W}_{1\mathcal A}\mathfrak W^{\mathcal A}_2-\bar{\mathfrak W}_{2\mathcal A}\mathfrak W^{\mathcal A}_1).
\end{array}
\end{equation}

Supertwistors are characterized by the following OPEs
\begin{equation}
\mathfrak Z^{\mathcal A}_1(z_1)\bar{\mathfrak Z}_{2\mathcal B}(z_2)\sim\frac{ic\alpha'}{z_{12}}\delta^{\mathcal A}_{\mathcal B},\quad\mathfrak Z^{\mathcal A}_2(z_1)\bar{\mathfrak Z}_{1\mathcal B}(z_2)\sim\frac{ic\alpha'}{z_{12}}\delta^{\mathcal A}_{\mathcal B}
\end{equation}
and
\begin{equation}
\mathfrak W^{\mathcal A}_1(\bar z_1)\bar{\mathfrak W}_{2\mathcal B}(\bar z_2)\sim\frac{ic\alpha'}{\bar z_{12}}\delta^{\mathcal A}_{\mathcal B},\quad\mathfrak W^{\mathcal A}_2(\bar z_1)\bar{\mathfrak W}_{1\mathcal B}(\bar z_2)\sim\frac{ic\alpha'}{\bar z_{12}}\delta^{\mathcal A}_{\mathcal B}.
\end{equation}
So that Virasoro and $GL(1,\mathbb C)$ generators of both sectors satisfy the OPEs
\begin{equation}
\begin{array}{c}
\psi_{\mathfrak Z}(z_1)\psi_{\mathfrak Z}(z_2)\sim\frac{2(N-4)}{z^2_{12}},\quad\chi_{\mathfrak Z}(z_1)\chi_{\mathfrak Z}(z_2)\sim\frac{2(N-4)}{z^2_{12}},\\[0.2cm]
T_{\mathfrak Z}(z_1)\psi_{\mathfrak Z}(z_2)\sim\frac{\psi_{\mathfrak Z}(z_2)}{z^2_{12}}+\frac{\partial\psi_{\mathfrak Z}}{z_{12}},\quad T_{\mathfrak Z}(z_1)\chi_{\mathfrak Z}(z_2)\sim\frac{\chi_{\mathfrak Z}(z_2)}{z^2_{12}}+\frac{\partial\chi_{\mathfrak Z}}{z_{12}},\\[0.2cm]
T_{\mathfrak Z}(z_1)T_{\mathfrak Z}(z_2)\sim\frac{2(N-4)}{2z^4_{12}}+\frac{2T_{\mathfrak Z}(z_2)}{z^2_{12}}+\frac{\partial T_{\mathfrak Z}}{z_{12}}
\end{array}
\end{equation}
and
\begin{equation}
\begin{array}{c}
\psi_{\mathfrak W}(\bar z_1)\psi_{\mathfrak W}(\bar z_2)\sim\frac{2(N'-4)}{\bar z^2_{12}},\quad\chi_{\mathfrak W}(\bar z_1)\chi_{\mathfrak W}(\bar z_2)\sim\frac{2(N'-4)}{\bar z^2_{12}},\\[0.2cm]
T_{\mathfrak W}(\bar z_1)\psi_{\mathfrak W}(\bar z_2)\sim\frac{\psi_{\mathfrak W}(\bar z_2)}{\bar z^2_{12}}+\frac{\partial\psi_{\mathfrak W}}{\bar z_{12}},\quad T_{\mathfrak W}(\bar z_1)\chi_{\mathfrak W}(\bar z_2)\sim\frac{\chi_{\mathfrak W}(\bar z_2)}{\bar z^2_{12}}+\frac{\partial\chi_{\mathfrak W}}{\bar z_{12}},\\[0.2cm]
T_{\mathfrak W}(\bar z_1)T_{\mathfrak W}(\bar z_2)\sim\frac{2(N'-4)}{2\bar z^4_{12}}+\frac{2T_{\mathfrak W}(\bar z_2)}{\bar z^2_{12}}+\frac{\partial T_{\mathfrak W}}{\bar z_{12}}.
\end{array}
\end{equation}
with the central charges $C=2(N-4)$, $\tilde C=2(N'-4)$. In the open string sector it should be $N=N'$ so that $C=\tilde C$ and the boundary conditions can be chosen as $\mathfrak Z_{1,2}^{\mathcal A}|_{\partial\rm M^2}=\mathfrak W_{1,2}^{\mathcal A}|_{\partial\rm M^2}$, $\bar{\mathfrak Z}_{1,2\mathcal A}|_{\partial\rm M^2}=\bar{\mathfrak W}_{1,2\mathcal A}|_{\partial\rm M^2}$.

For the model under consideration the Virasoro generator of the holomorphic sector $T_{\mathfrak Z}$ can be twisted by the $U(1)_{\mathfrak Z}$ current
\begin{equation}
T_{\mathfrak Z d}(z)=T_{\mathfrak Z}+d\partial\chi_{\mathfrak Z}
\end{equation}
that shifts conformal weights of supertwistors as
\begin{equation}
T_{\mathfrak Z d}(z_1)\mathfrak Z^{\mathcal A}_1(z_2)\sim\frac{(1+d)\mathfrak Z^{\mathcal A}_1(z_2)}{2z^2_{12}}+\frac{\partial\mathfrak Z^{\mathcal A}_1}{z_{12}},\quad
T_{\mathcal Z d}(z_1)\mathfrak Z^{\mathcal A}_2(z_2)\sim\frac{(1-d)\mathfrak Z^{\mathcal A}_2(z_2)}{2z^2_{12}}+\frac{\partial\mathfrak Z^{\mathcal A}_2}{z_{12}}
\end{equation}
and the value of the central charge changes to $C_d=2(N-4)(1-3d^2)$. Again for the case of $N=4$ supertwistors their  contribution to conformal and $GL(1,\mathbb C)$ anomalies vanishes regardless of the twisting. However, to cancel conformal anomaly contribution of the ghosts corresponding to gauge symmetries it is necessary to add extra matter with $C=30$. The same conclusions are valid for the antiholomorphic sector variables.

\section{Conclusion and discussion}

There has been explored Hamiltonian mechanics of the
$\kappa-$symmetry gauge-fixed Lorentz-harmonic $D=4$ $N=1,2$
superstring in the supertwistor formulation. Superstring action
has been shown to be characterized by the primary and secondary
constraints that were identified and classified according to the
Dirac prescription. The superstring action is invariant under the
set of gauge transformations generated by the first-class
constraints. These include world-sheet reparametrizations, local
$SO(1,1)$ rotations, as well as, Weyl dilations and two
independent $U(1)$ rotations of the supertwistor components provided the components of
unnormalized Newman-Penrose dyad are used as auxiliary variables. The second-class
constraints of the model could be taken into account by
introducing the D.B. We have started the realization of this
approach by constructing the D.B. depending on the part of the
second-class constraints that reproduce known permutation
relations of the twistor algebra and evaluated the D.B. of the
remaining second- and first-class constraints. Alternative
approach to the treatment of the second-class constraints based on
their conversion \cite{Fad}, \cite{BF} to the effective
first-class constraints also deserves further study.

Using methods of $2d$ CFT there have been examined quantum realizations of the reparametrization and $U(1)$ symmetries. There have been presented the OPEs of the Virasoro and $U(1)$ generators that determine the quantum algebra of the superstring and the values of conformal and $U(1)$ anomalies were obtained. Considered here $N=1$ supertwistor formulation of the $D=4$ Lorentz-harmonic superstring appears to be incomplete in the sense that in order to cancel anomalies it is necessary to add extra matter degrees of freedom, whose properties, as well as, the quantum states of the resulting model require further investigation. They could be e.g. degrees of freedom generating nonAbelian gauge symmetry.

The last part of the paper has been devoted to the study of conceivable generalizations. It is possible to consider string models based on $N$-extended supertwistors. Though they can not be related to the $D=4$ Green-Schwarz superstring they could be related to reductions of higher dimensional superstrings. Of particular interest is the $N=4$ case, where as the result of cancellation between contributions of commuting and anticommuting components, $U(1)$ and conformal anomalies do not receive contributions from the supertwistors. This $N=4$ superstring model, however, is different from twistor superstrings \cite{Berkovits}, \cite{BerkovitsMotl} as the action is invariant only under the $U(1)$ gauge symmetry rather than $GL(1,\mathbb C)$ necessary for the supertwistor components to be identified with the homogeneous coordinates of the projective twistor space $\mathbb{CP}^{(3|4)}$. So other considered generalization is to add extra pair of supertwistors that allows to gauge the whole $GL(1,\mathbb C)$ and also to consider different values of conformal weights by twisting stress-energy tensor. The question whether such superstring is related to that of \cite{Berkovits}, \cite{BerkovitsMotl}, defined initially for real supertwistors corresponding to space-time of signature $(2,2)$, as well as, other features of this model require further study.

Finally let us note that one of the issues of the twistor superstring approach to the gauge/string correspondence is to find an extension to $D=4$ gauge theories with $N<4$ supersymmetry and nonconformal (in the $4d$ sense) theories. For the progress towards the solution see e.g. \cite{Popov}-\cite{Hull}. For considered here models the number of supersymmetries can easily be varied and the superconformal symmetry can be broken down to superPoincare one by introducing the term of the form $u^\alpha v_\alpha\bar u^{\dot\alpha}\bar v_{\dot\alpha}$ that coincides with the mass squared operator in the twistor theory.

\section{Acknowledgements}

The author is grateful to A.A.~Zheltukhin for the valuable discussions.

\end{document}